# A metal-insulator transition of the buried MnO$_2$ monolayer in complex oxide heterostructure

Heng-Jui Liu[1,2,†], Jheng-Cyuan Lin[3,†], Yue-Wen Fang[4], Jing-Ching Wang[5], Bo-Chao Huang[3], Xiang Gao[6], Rong Huang[4,6], Philip R. Dean[7], Peter D. Hatton[7], Yi-Ying Chin[8], Hong-Ji Lin[8], Chien-Te Chen[8], Yuichi Ikuhara[6,9], Ya-Ping Chiu[2,5], Chia-Seng Chang[3], Chun-Gang Duan[4], Qing He[7*], and Ying-Hao Chu[1,3*]

[1] *Department of Materials Science and Engineering, National Chiao Tung University, Hsinchu 30010, Taiwan*

[2] *Department of Physics, National Taiwan Normal University, Taipei 116, Taiwan*

[3] *Institute of Physics, Academia Sinica, Taipei 11529, Taiwan*

[4] *Key Laboratory of Polar Materials and Devices, Ministry of Education, East China Normal University, Shanghai 200241, China*

[5] *Department of Physics, National Sun Yat-sen University, Kaohsiung, 804, Taiwan*

[6] *Nanostructures Research Laboratory, Japan Fine Ceramics Center, Nagoya 456-8587, Japan*

[7] *Department of Physics, Durham University, Durham DH1 3LE, United Kingdom*

[8] *National Synchrotron Radiation Research Center, Hsinchu 30076, Taiwan*

[9] *Institute of Engineering Innovation, The University of Tokyo, Tokyo 113-8656, Japan*

**Functionalities in crystalline materials are determined by 3-dimensional collective interactions of atoms. The confinement of dimensionality in condensed matter provides an exotic research direction to understand the interaction of atoms, thus can be used to tailor or create new functionalities in material systems. In this study, a 2-dimensional transition metal oxide monolayer is constructed inside complex oxide heterostructures based on the theoretical predictions. The electrostatic boundary conditions of oxide monolayer in the heterostructure is carefully designed to tune the chemical, electronic, and magnetic states of oxide monolayer. The challenge of characterizing such an oxide monolayer is overcome by a combination of transmission electron microscopy, x-ray absorption spectroscopy, cross-sectional scanning tunneling microscopy, and electrical transport measurements. An intriguing metal-insulator transition associated with a magnetic transition is discovered in the MnO$_2$ monolayer. This study paves a new route to understand the confinement of dimensionality and explore new intriguing phenomena in condensed matters.**

Due to the interplay of lattice, charge, orbital, and spin degrees of freedom, strongly correlated electrons in complex oxides generate a rich spectrum of competing phases and emergent physics[1, 2]. Recently, extensive studies suggest that complex oxide interfaces provide a powerful route to manipulate these degrees of freedom and offer new possibilities for next generation devices, thus create a new playground for investigating novel physics and the emergence of fascinating states in condensed matter[3, 4]. In 2004, a two-dimensional (2D) electron gas was discovered at the $LaAlO_3/SrTiO_3$ (LAO/STO) heterointerface[5]. Since then, the LAO/STO interface shows ultrahigh mobility, interface superconductivity[6], magnetoresistance[7], and multiple tunable properties[8, 9, 10]. On the other hand, the discovery of free-standing 2D materials has inspired the research for exploring new low-dimensional materials[11, 12, 13]. Pioneered by graphene[14], these 2D materials exhibit aboundant unusual physical phenomena that is undiscovered in bulk forms. The confinement of charge and heat transport at such ultrathin planes offers possibilities to overcome the bottleneck of current devices. Furthermore, researchers have speculated that superior properties such as high temperature superconductivity and high mobility electronic transport can occur at an oxide plane. For example, a single $CuO_2$ plane with a superconducting temperature ranging from 15 K-50 K could be procured at the interface between metallic $La_{1.55}Sr_{0.45}CuO_4$ and insulating $La_2CuO_4$[15]. Moreover, in some systems, the properties have been suggested to dominate by their 2D structural units, *i.e.* the copper oxide plane in $YBa_2Cu_3O_{7-x}$[16], or the ruthenium oxide plane in $Sr_2RuO_4$ are responsible for their superconductivity[17]. In this study, we explore the heterostructures that combine the merits of two intriguing systems, the state-of-the-art of complex oxide interfaces and 2D materials. 2D oxide monolayer will be sandwiched between complex oxides. In such heterostructures, the intrinsic properties of 2D oxide monolayer can provide a basic understanding of the physics on the dimensionality-confined correlated electrons;

while various oxide heterostructure will be employed to modulate the properties of 2D oxide monolayer. This study delivers a generic approach to study the dimensional confinement of correlated electrons and provides a direction to design new electronic devices.

The concept is illustrated in Fig. 1. A transition metal oxide monolayer is buried in complex oxide heterostructures. In a model system of $MnO_2$ monolayer, three heterostructures are studied in parallel. In case 1, the $MnO_2$ monolayer is embedded into a STO crystal, such that the top and bottom neighboring layers of the $MnO_2$

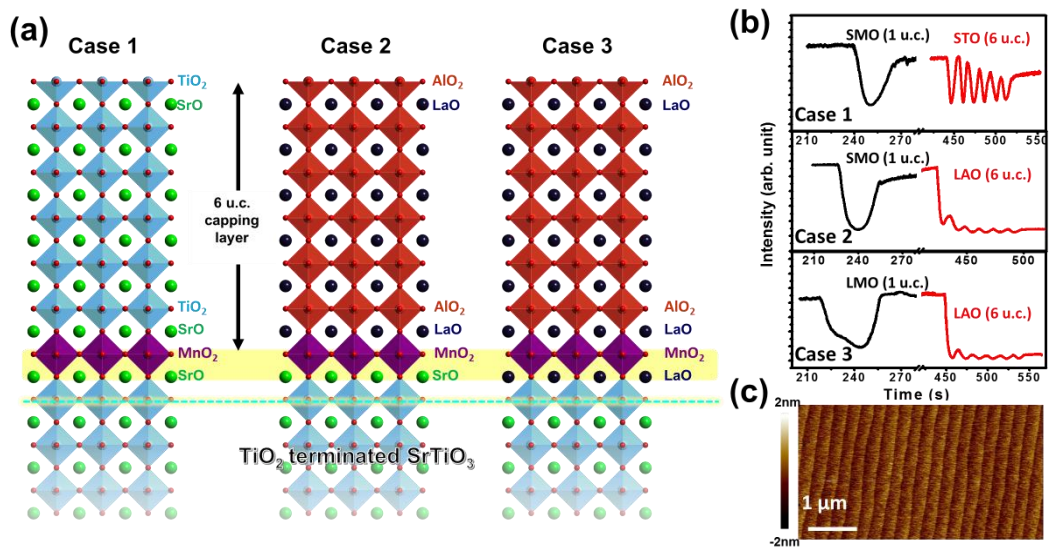

Figure 1 **a,** Schematics of the designed $MnO_2$ monolayers with different electrostatic boundary conditions. case 1: $SrO^0/MnO_2/SrO^0$, case 2: $LaO^+/MnO_2/SrO^0$, case 3: $LaO^+/MnO_2/LaO^+$. **b,** *Real time* monitoring of RHEED intensity for these three cases. Case 1: 1 unit cell of SMO grown on $TiO_2$ terminated STO capped with 6 unit cells of STO. Case 2: 1 unit cell of SMO grown on $TiO_2$ terminated STO capped with 6 unit cells of LAO. Case 3: 1 unit cell of LMO grown on $TiO_2$ terminated STO capped with

6 unit cells of LAO. **c,** The typical morphology of atomically flat surface for these three cases.

monolayer are both SrO. Because of the charge neutrality of SrO layers, the valence state of Mn ions is expected to be +4, in which their 3$d$ electrons fill up to the $t_{2g}$ orbitals. In case 2, the MnO$_2$ monolayer is inserted into a LAO/STO heterointerface. The top neighbor layer of the MnO$_2$ monolayer is LaO$^+$ and the bottom neighbor layer is SrO. The charge transfer is expected to drive the valence state of Mn ions between +3 and +4. Thus, the extra 3$d$ electrons start to fill into $e_g$ orbitals. The mixture of Mn$^{4+}$ and Mn$^{3+}$ ions in manganites is famous for exhibiting novel physical phenomena[18]. The design concept of case 3 is similar to case 2, but going further. In this structure, both the top and bottom neighbor layers of the MnO$_2$ monolayer are designed to be LaO$^+$. The valence state of Mn ions, in this case, should be pushed further towards +2, resulting in a mixture of multi-valence Mn ions. Because the charge transfer is expected to emerge and modify the electronic and magnetic structures of the MnO$_2$ monolayers, a question arises: what are the magnetic ground states of the MnO$_2$ monolayers in these cases?

|        | FM (eV)      | AFM (eV)     | FM-AFM (meV)/(Mn-Mn pair) |
|--------|--------------|--------------|---------------------------|
| Case 1 | -1126.468995 | -1126.546111 | 19.3                      |
| Case 2 | -1127.42961  | -1127.314838 | -28.6                     |
| Case 3 | -1139.0685   | -1138.6896   | -94.7                     |

Table 1. Results of total energy calculations using $U_{Mn}^{eff}$ = 4.5 eV. FM and AFM denote the ferromagnetic and antiferromagnetic configurations of the monolayer MnO$_2$ in each case.

To address this question, we employed the density functional theory (DFT) method to perform total-energy calculations for the heterostructures (see Methods: DFT calculation for details). The results of the respective total energies for each case in ferro/antiferro-magnetic configuration are listed in Tab. 1 using $U_{Mn}^{eff}$ = 4.5 eV. One can clearly see that Mn-Mn pair in case 1 exhibits antiferromagnetic coupling, whereas ferromagnetism is stabilized in the ground state for case 2 and case 3. Due to the significant difference in the magnetic ground state, a dramatic difference in electronic structures of these heterostructures can be expected. We then computed the electronic structures of the heterostructures and presented the obtained density of states (DOS) extracted from spin-polarized calculations with $U_{Mn}^{eff}$ = 4.5 eV in Fig. 2. From the total DOS, it is clear that case 1 is an insulator with an energy gap of at least 0.5 eV and case 3 is a metal with the Fermi level shifted to the conduction bands of both spin channels. However, case 2 is more complicated. It seems to be an insulator since the total DOS (middle left panel of Fig. 2) is analogous to case 1. However, from the projected DOS of the MnO$_2$ layer, as shown in the middle right panel of Fig. 2, one can see that case 2 indeed displays a half-metallicity with an energy gap of about 0.53 eV in the minority spin channel. It should be noted that the Fermi level shifts up gradually from case 1 to case 2 to case 3, indicating an increasing electronic occupation of Mn ions in these heterostructures. We hence conducted the analysis of Bader charge for Mn ions in each case, and found that the valence state of Mn ions meets the relationship of case 1 > case 2 > case 3. We futher examined the electronic charge by integrating the Mn 3$d$ orbital resolved density of states. The electronic charge of Mn in each case is 4.73, 4.80, and 4.83, respectively. These semi-quantitative results clearly show charge transfer really occurs, and the oxidization state of Mn in case 1 (3) is the biggest (smallest) one among the three cases.

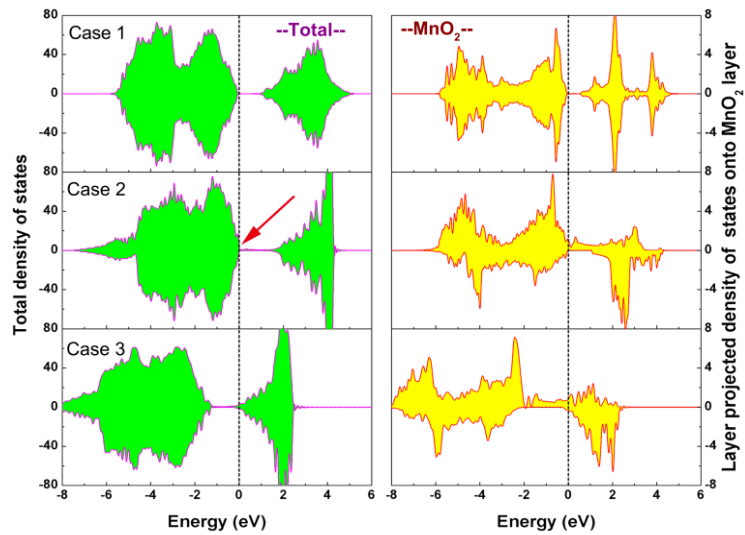

**Figure. 2 Density of states of the heterostructures.** The left panel shows the total DOS of each case and the right panel shows the layer projected DOS for MnO$_2$ layer in corresponding case. Zero is the reference for the Fermi level. The DOS at the Fermi level in case 2 pointed by a red arrow must be given a particular attention because it actually displays half-metallic conductivity in DFT calculations which is shown in the right panel.

Inspired by the theoretical prediction, we turned our focus to the sample preparation. Single TiO$_2$-terminated (001) STO substrates[19] were used because it is a crucial requirement for obtaining the layer-by-layer growth and precisely controlling surface termination of subsequent layers,. To *in-situ* monitor the growth mode, pulsed laser deposition (PLD) equipped with a high-pressure reflection high energy electron diffraction (RHEED) facility has been adopted. The intensity variation of the reflected electron beam was monitored. When a full period of oscillation finished by first decreasing to the minimum and then coming back to the maximum indicates the completion of one atomic layer[20, 21]. Thus, one unit cell of SrMnO$_3$ (SMO) or LaMnO$_3$ (LMO) to form the MnO$_2$ monolayer can be grow and then covered with different capping layers such as neutral STO or polar LAO films. In case 1, while one unit cell of SMO layer has been deposited onto TiO$_2$ terminated STO substrate, the SrO plane naturally connects to the TiO$_2$ termination, thus making the MnO$_2$ plane as the new termination on the surface for keeping the structure continuity. Thereafter, this new termination has been covered by six unit cell STO film to achieve the architecture of case 1. In case 2, the one unit cell of SMO layer has been kept but the capping layer is replaced by six unit cells of LAO. In case 3, unlike the $MnO_2$ monolayers possessing electronic neutrality in case 1, the polar $MnO_2$ monolayer in case 3 is developed by growing one unit cell of LMO layer, and then covered by six unit cells of LAO. In order to obtain the desired architecture, the deposition process was recorded *in-situ* by the RHEED, where SMO or LMO part (in black) and STO or LAO (in red) exhibited one and six complete oscillations, respectively, as shown in Fig. 1(b). After the growth, the surfaces of all samples remained atomically smooth. A typical image captured by atomic force microscopy (AFM) is shown in Fig. 1(c).

(a)

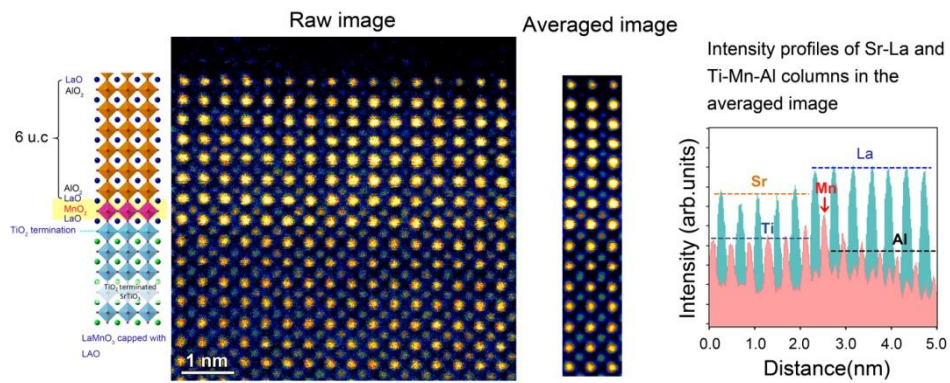

(b)

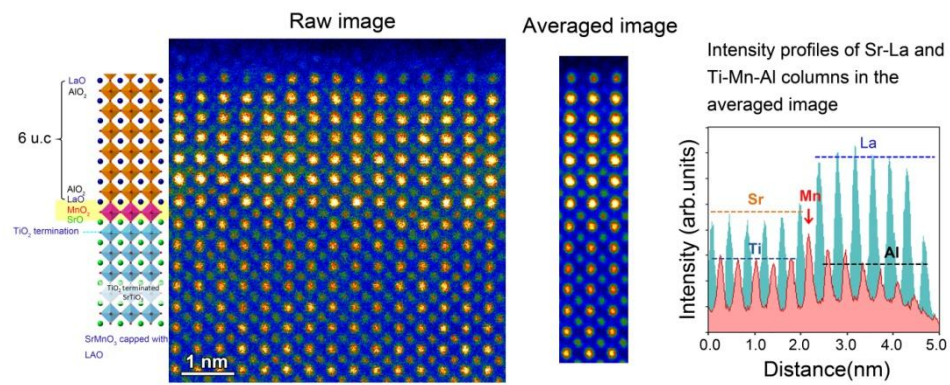

(c)

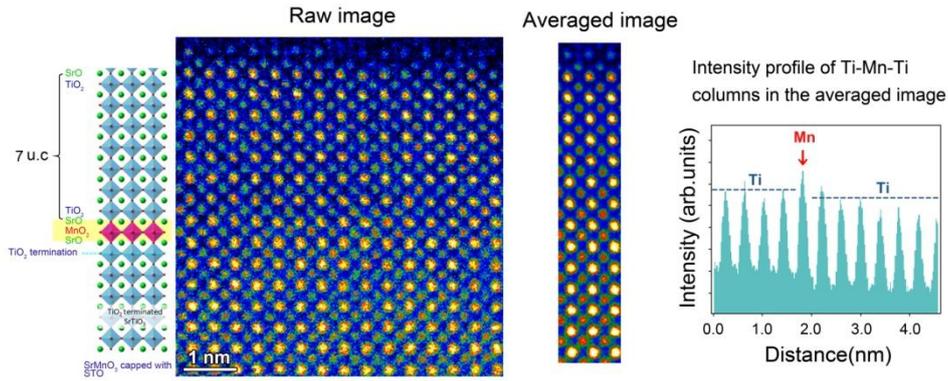

Figure 3. Typical HAADF-STEM images of the three $MnO_2$ monolayer samples. **a,** Raw image (left), averaged image (middle) and the intensity profiles of the Sr-La and Ti-Mn-Al columns (right) in the averaged image of the LMO monolayer capped with 6 unit cells of LAO observed along the [010] zone axis of STO substrate. **b,** Raw image (left), averaged image (middle) and the intensity profiles of the Sr-La and Ti-Mn-Al columns (right) in the averaged image of the SMO monolayer capped with 6 unit cells of LAO observed along the [010] zone axis of STO substrate. **c,** Raw image (left), averaged image (middle) and the intensity profiles of the Ti-Mn columns (right) in the averaged image of the SMO monolayer capped with 6 unit cells of STO observed along the [010] zone axis of STO substrate.

To confirm the fabrication of the desirable heterostructures, high-angle annular dark-field scanning transmission electron microscopy (HAADF-STEM) was carried out. Fig. 3(a) shows a typical HAADF-STEM image of case 3 along the [010] direction of the STO substrate. The atomic columns of A sites (Sr and La) and B sites (Ti, Mn, and Al) can be seen in the raw image, as shown in the left panel of the Fig. 3(a). Because the atomic number of La ($Z = 57$) is much larger than that of Sr ($Z = 38$), both atomic columns can be distinguished unambiguously. The columns of La exhibit brighter contrast than those of Sr. However, the columns of the Mn monolayer are very difficult to be differentiated from those of Ti due to a small difference in their atomic numbers of Mn (Z= 25) and Ti (Z= 22). The difference in the contrast between Mn and Ti is almost comparable to the noise level in the raw image. To improve the signal-to-noise ratio of the raw image, an average process has been performed, as shown in the middle of the Fig. 3(a). In the averaged image, the heavier Mn columns exhibit relatively brighter contrast than the lighter Ti columns, as revealed in the intensity profile of the Ti-Mn-Al columns, shown in the right panel of the Fig. 3(a). By comparing the intensity profiles of the Sr-La columns and the Ti-Mn-Al columns, the existence of the $MnO_2$ monolayer in case 3 can be confirmed. With the same method, the $MnO_2$ monolayers in case 2 and case 1 are also characterized, as shown in Figs. 3(b) and (c), respectively. Especially, the different sequence of the $MnO_2$ monolayer and the LaO layer in case 3 and case 2 was clearly revealed in the right panels of the Fig. 3(a) and 3(b).

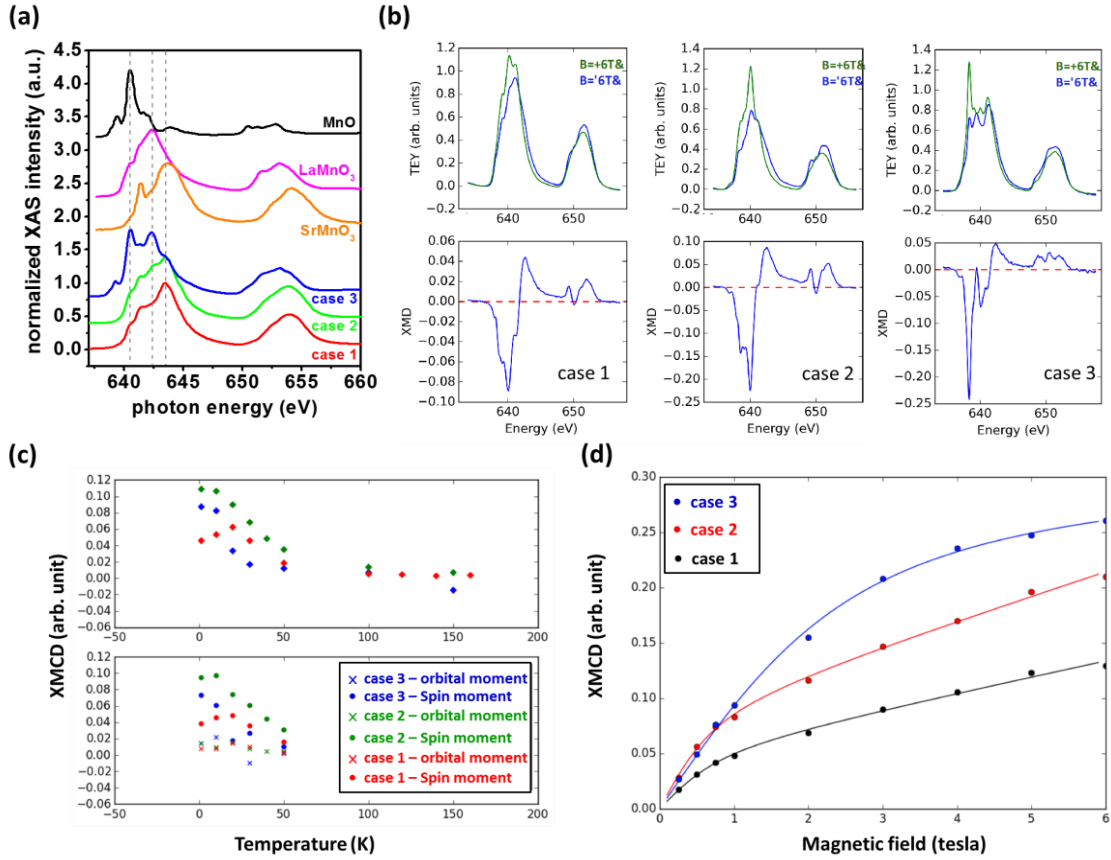

Figure 4: **a,** The XAS spectra across the Mn *L*-edges taken from the MnO monolayers with various electrostatic boundary conditions. **b,** The XMCD spectra across the Mn *L*-edges taken from the $MnO_2$ monolayers with various electrostatic boundary conditions. **c,** The extracted orbital and spin moments of the $MnO_2$ monolayers based on sum rules as a function of temperature with various electrostatic boundary conditions. (d) The XMCD signal of the $MnO_2$ monolayers as a function of magnetic field with various electrostatic boundary conditions.

Soft x-ray absorption based techniques are ideal tools to provide chemical, electronic, and magnetic information of these $MnO_2$ monolayers, as they are inherently element-specific, allowing the characterization of the valence states and the symmetry of individual lattice sites[22, 23, 24]. Fig. 4(a) shows the x-ray absorption spectra (XAS) across the Mn *L*-edges taken from the heterostructures. The changes of the XAS among various cases provide a direct evidence of the change in the valence state of Mn ions. For comparison, three reference spectra for $Mn^{4+}$ (SMO), $Mn^{3+}$ (LMO), and $Mn^{2+}$ (MnO) ions are added in the upper part of Fig. 4(a). Three dashed lines in this figure indicate the peak position of signature absorption for $Mn^{4+}$, $Mn^{3+}$, and $Mn^{2+}$ ions. The corresponding signature peaks of the $MnO_2$ monolayers for the three cases can be observed in the XAS, suggesting a systematic change of the valence state of Mn ions. In case 1, the valence state of Mn ions is dominated by $Mn^{4+}$ ions. In case 2, a mixed valence state of $Mn^{3+}$ and $Mn^{4+}$ was detected. Further, in case 3, we discovered the valence state of Mn ions being dominated by a mixture of $Mn^{+2}$ and $Mn^{3+}$. As suggested by the DFT calculation, we further investigated the magnetic states of these $MnO_2$ monolayers. Thus, x-ray magnetic circular dichroism (XMCD) measurements across the Mn $L$-edges as a function of magnetic field and temperature were employed to probe the magnetic phases and possible phase transitions. Fig. 4(b) shows the XMCD spectra of the heterostructures measured at ~3K with an applied magnetic field of ~6T. Significant differences in both magnitudes and line shapes of XMCD spectra in these $MnO_2$ monolayer were found. The smallest XMCD was detected in case 1, while case 3 showed the largest XMCD, indicating a stronger magnetic response of case 3. In order to obtain quantitative magnetometry, sum rules have been applied to the XAS-XMCD spectra to extract the spin moment and orbital moment of Mn ions shown in Fig. 4(c). The magnitude of orbital moment in these samples was found to be very close and show very little temperature dependence, which is consistent with the expectation. With the largest magnitude, the spin moment of case 3 shows a clear temperature dependence with a transition around 60K. Similar behavior can be found in other two cases. The magnetic-filed dependent XMCD curves are shown in Fig. 4(d), a curve fitting based on superparamagnetism can be used to fit the behavior. These results suggest the systems possess high magnetic susceptibility, a nature of 2D magnetism. Based on the

XAS-XMCD results, the electronic and magnetic structures of the MnO$_2$ monolayers can be manipulated by a careful control of the electrostatic configurations.

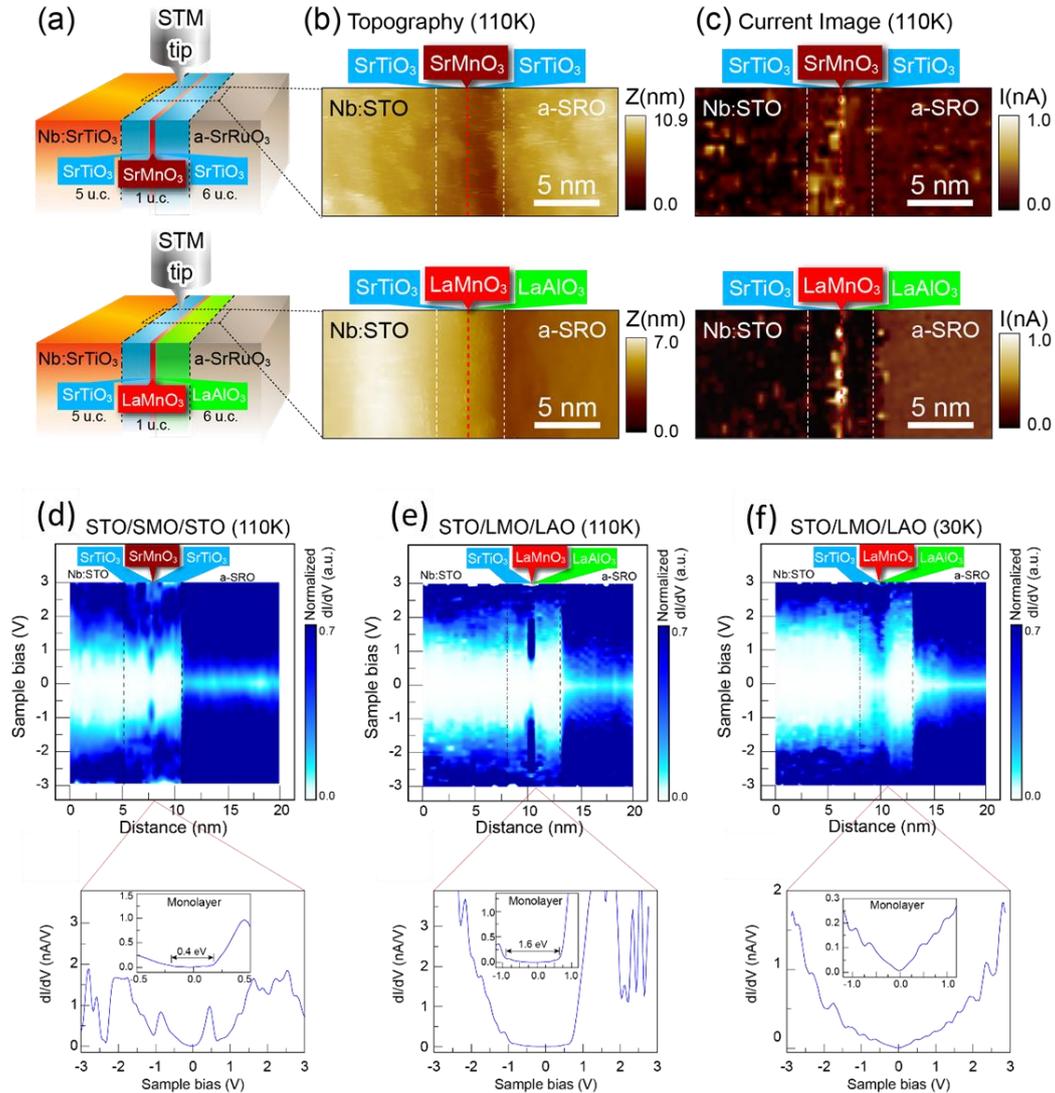

Figure 5: **a,** Schematics of cross-sectional STM employed to investigate the properties of the MnO$_2$ monolayers. **b,** Typical cross-sectional topography images of the case1 and case 3 at sample bias of -2.0 V. **c,** STM/S spectroscopic current images of the case1 and case 3 at sample bias of -2.0 V. The red dashed lines in case 1 and case 3 can be assigned as the position of the MnO$_2$ monolayers, which show brighter contrast compared to other region. **d,** Band alignment and the corresponding atomic-scale evolution of electronic properties across the MnO$_2$ monolayers in case 1 are measured at 110 K. **e-f,** The same analyses of case 3 are measured at 110 K and 30 K, respectively.

In order to probe the electronic structure of these $MnO_2$ monolayers in real space, the cross-sectional scanning tunneling microscopy and spectroscopy (XSTM/XSTS) was carried out on the surface of cleaved samples in an ultrahigh vacuum (UHV) chamber[25]. The schematic of XSTM measurements of case 1 and case 3 are illustrated in Fig. 5(a). This technique can provide the information of local density of states (LDOS) spatially with atomic-level resolution. Fig. 5(b) shows the typical STM topography across the heterointerfaces of case 1 and case 3 with applied voltage of -2.0 V. Various regions can be distinguished based on the specific spectra among differnt layers. The corresponding spectroscopic results for the heterostructures of case 1 and case 3 are shown in Fig. 5(c). Compared to the *local* electronic information of case 1 under +1.5 V at 110K, a dramatically enhanced electronic signal was observed at the interface of case 3, revealing the critical signature of the electronic structures at the $MnO_2$ monolayer of case 3. Furthermore, the evolution of the band alignment across heterointerfaces of case 1 and case 3 was quantitatively built and mapped shown in Fig. 5(d) and Fig. 5(e) to illustrate the trivial discrepancy in the electronic structure. Based on the spectroscopic data, the energy gap of the $MnO_2$ monolayer in case 1 is about 0.4 eV, whereas the energy gap of the $MnO_2$ monolayer is about 1.6 eV in case 3, showing that the $MnO_2$ monolayers in these cases are semiconducting. Based on the dI/dV curves of the $MnO_2$ monolayers at 110K, we found that the density of states of the $MnO_2$ monolayer in case 1 is much smaller than that of case 3, suggesting that the electronic conductivity of the $MnO_2$ monolayer in case 3 is much higher than that in case 1. In order to study the phase transition of the $MnO_2$ monolayer, the band alignments of the heterostructure in case 3 at various temperatures were measured and mapped in Fig. 5(e) and 5(f). A significant difference between the energy gap of the $MnO_2$ monolayer at 30 K (about zero) and at 110K (about 1.6 eV) was observed. The results suggest that the $MnO_2$ monolayer in case 3 is in the metallic state at low temperatures but in the semiconducting state at high temperatures, revealing a metal-insulator transition. This transition temperature is consistent with the one found in the XMCD measurements.

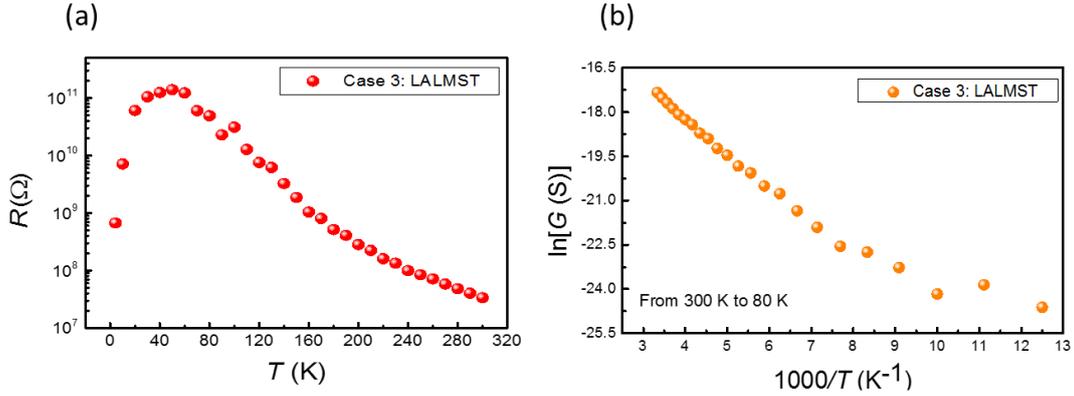

Figure 6 **a,** A metal-insulator transition with $T_C$ around 70 K found in case 3 based on the result of resistance vs. temperature. **b,** lnG as a function of temperature for the case 3 form 60 K to 300 K, very well describing Arrhenius equation G(T) within this temperature range.

To confirm the results of XAS-XMCD and XSTM measurements, we further carried out macroscopic electronic transport measurements. Due to the nature of high insulating states and the limitation of the instrument, the transport data are excluded for case 1 and case 2. For case 3, a standard two-terminal transport measurement over the temperature range from 4.2 K to 300 K was performed. The resistance versus temperature of case 3 is shown in Fig. 6(a). The resistance decreases with increasing temperature in the temperature range from 60 K to 300 K, exhibiting a semiconducting behavior. Such a semiconducting behavior can be described by several models such as localization and thermal excitation. Fig. 6(b) shows that the logarithm of the conductance (*G*) of case 3 is linearly proportional to the temperature. Obviously, the transport behavior in the insulating regime of case 3 follows Arrhenius equation *G(T)*

$=G_0 \exp(E_A/K_B T)$ very well ($G_0$ is a prefactor of conductance). The thermal activation energy $E_A$ to be about 170 meV was extracted. More importantly, a clear metal-insulator transition was observed at ~60 K (Fig. 6(a)), which is consistent with the observations from XSTM and XMCD measurements. In this case, the theoretical calculations show that the ground state is metal, which agrees with the experimental observation at low temperature. However, when the temperature reaches above $T_C$, i.e. the system enters a paramagnetic state, the system should behave more like an insulator attributed to the enhanced random spin scattering in paramagnetic state with local magnetic dipoles[26,27].

In summary, we have successfully demonstrated that a sandwiched 2-dimensional metal oxide monolayer can be fabricated by a precise control of the growth in atomic scale. The top capping layers, STO and LAO, provide the different electrostatic boundary conditions, which obviously influence the chemical, electronic, and magnetic states of the buried $MnO_2$ monolayers. Interestingly, the $MnO_2$ monolayer in case 3 is predicted to be a metal, which agrees with the experimental observation at low temperature. Moreover, an intriguing metal-insulator transition has been found to correlate with a magnetic transition in case 3. Therefore, the achievement of this study not only proposes a new concept to study the confinement of strongly correlated electrons in low dimension, but also offers the potential in modulating relevant novel phenomena in the domain of 2D strongly correlated electron systems.

**Method:**

**Sample preparation**

All samples were fabricated from stoichiometric targets (SMO, LMO, STO, and LAO) on the $TiO_2$-terminated STO (100) substrates using pulsed laser deposition equipping with a RHEED. During the deposition, the substrate temperature was maintained at 700 °C. The oxygen pressure was kept at 100 mtorr for the deposition of SMO, LMO, and STO, and ~$5 \times 10^{-5}$ torr for the LAO growth, respectively. After the deposition, all the films were post-annealed at the oxygen pressure of approximately 300 torr for 30min, and then cooled down to room temperature.

**DFT calculation**

In the first-principles calculation within the framework of density-functional theory, the exchange-correlation potential is treated in local density approximation (LDA) and local density approximation plus Hubbard $U$ (LDA+$U$) while the projected augmented wave method is used with a plane wave basis set as implemented in the Vienna *ab initio* simulation package (VASP)[28, 29]. A kinetic energy cut off is set to be 500 eV for the plane wave basis and the Brillouin zone integration for self-consistent field calculation is carried out using Monkhorst-Pack grid of $5 \times 5 \times 1$ k points for heterostructures and $8 \times 8 \times 8$ k points grid for bulk STO and LAO in combination with the tetrahedron method. The electronic self-consistency convergence is set to be $10^{-5}$ eV for all calculations. The optimized bulk lattice constants by LDA are 3.874 and 3.742 Å for STO and LAO, respectively, which are slightly underestimated with respect to their experimental values (3.905 and 3.789 Å) and are in agreement with Han's theoretical predictions[30]. In the optimization for the three cases, we fixed the in-plane lattice constant of the supercells at the relaxed lattice constant of bulk STO and performed relaxation of all the coordinates of atomic positions along the *c*-direction until the Hellmann-Feynman forces on each atom were less than 1 meV/Å. On account of the on-site electron-electron interactions in the localized 3d orbitals for Mn, we used LDA+U method introduced by Dudarev et al.[31] to get the Hamiltonian including screened Coulomb interactions. In order to investigate the effect on the Hamiltonian introduced by Hubbard correction, we carefully performed the calculations with the values $U_{Mn}^{eff} = 2, 4, 4.5$ and 7 eV, respectively. $\sqrt{2} \times \sqrt{2} \times 13$ geometrical models were constructed to simulate the heterostructures. The qulitative results in our study are verified by both slab and superlattice models. Notably the slab model were simulated by adding imaginative vacuum thickness of 15 Å on top of capping layers and the dipole correction was included in calculations to eliminate the spurious field.

**Structural analysis by HAADF-STEM**

A Cs-corrected STEM (JEM-2100F, JEOL, Co., Tokyo, Japan) operated at 200 kV and equipped with a spherical aberration corrector (CEOS Gmbh, Heidelberg, Germany) were performed with a minimum probe of about 1 Å in diameter. The probe convergence angle and the detection angle were 25 mrad and 92–228 mrad, respectively, during HAADF imaging. The low loss EELS analyses were carried out using a Cs-corrected STEM (JEM-ARM200F, JEOL, Co., Tokyo, Japan) operated at 200 kV and equipped with a spherical-aberattion corrector (CEOS GmbH, Heidelberg, Germany), a Gatan Image Filter (GIF) and a Wien-filter type monochromator. An entrance aperture of 2.5 mm was used and the energy resolution of 0.15 eV was determined by measuring the full-width half-maximum (FWHM) of the zero loss peak.

**Magnetism study by soft X-ray absorption techniques.**

The electronic structure and origin of magnetism of the $MnO_2$ monolayers were studied using soft X-ray absorption spectra (XAS) and X-ray magnetic circular dichroism (XMCD) at the Dragon beamline in NSRRC. The temperature and magnetic field dependent spectra were recorded at the Mn $L_{2,3}$ edges (630 - 660eV) with an energy resolution of 0.25 eV by shining Left and right circularly polarized X-rays on the samples. The measured temperature ranged from 3K to room temperature. The applied magnetic field were performed from 0.2 T to 6 T.

**Cross-section scanning tunneling microscopy and spectroscopy (XSTM/STS)**

Before the STM measurements, amorphous SrRuO$_3$ (SRO) capping layers (>500 nm) were deposited on top of all samples to prevent tip crash during the measurements. These samples were cleaved *in situ* in an UHV chamber whose base pressure was approximately $5 \times 10^{-11}$ torr. The cross-sectional STM topography images was obtained using constant current mode at -2.0 V. The corresponding atomically resolved spectroscopic results with the spatial resolution of 0.4 nm at interfaces were recorded with the first derivative of tunneling current over tip-sample voltage (or differential conductivity), dI/dV, at temperature of 30 K and 110K.

**Electric transport measurements**

The typical transport measurement was performed on Cryogen-free 4 K cryogenic probe station (Lake Shore Model CRX-4K) with high resistance meter (Keithley Model 6517A) for the three cases samples. The conductance results which are shown in this paper are from inverse resistance, G = 1/R and the resistance was obtained under the DC electric bias 1 V.

**Acknowledgements**

The authors thank Dr. Takeharu Kato and Ryuji Yoshida at the Japan Fine Ceramics Center (JFCC) for their help in preparing TEM samples. The authors thank Prof. Wen-Bin Jian's laboratory for their assistance in the electric transport measurement. The authors acknowledge the support of the Ministry of Science and Technology under Grant No. MOST 103-2119-M-009-003-MY3. This work was also supported in part by the National Key Project for Basic Research of China (No.2014CB921104 and 2013CB922301), the NSF of China (No. 61125403). It is also supported in part by Ministry of Science and Technology of Taiwan (Grants 103-2745-M-002-004-ASP). Computations in this work were performed at ECNU computing center and Chinese Tianhe-1A system at the National Supercomputer Center.

**Author information**

*E-mail: yhc@nctu.edu.tw and qing.he@durham.ac.uk. Address: Room 709, Engineering Building VI, 1001 University Road, Hsinchu 30010, Taiwan. Phone: +886-972-781-386 Author

**Contributions**

†These authors contributed equally to this work.